\documentclass[conference, letterpaper]{IEEEtran}

\usepackage{nameref} 
\usepackage{float} 
\usepackage[pdftex]{graphicx} 
\usepackage{indentfirst} 
\usepackage{tabu} 
\usepackage{epstopdf} 
\usepackage{url}
\usepackage{hyperref} 
\title{Making Defeating CAPTCHAs Harder for Bots}
\author{ Nasser Mohammed Al-Fannah \\ Information Security Group \\ Royal Holloway, University of London \\ \href{mailto:nasser@alfannah.com}{nasser@alfannah.com}
\\~\\11 October 2016
}

\begin{document}
	\maketitle
	\begin{abstract}
		For a number of years, many websites have used CAPTCHAs to filter out interactions by bots.  However, attackers have found ways to circumvent CAPTCHAs by programming bots to solve or bypass them, or even relay them for humans to solve.  In order to reduce the chances of success of such attacks, CAPTCHAs can be strengthened by the addition of certain safeguards.  In this paper, we discuss seven existing safeguards as well as five novel safeguards designed to make circumventing CAPTCHAs harder.  These safeguards are not mutually exclusive and can add multiple layers of protection to a CAPTCHA.  We further provide a high-level comparison of their effectiveness in addressing the threat posed by CAPTCHA-defeating techniques.  In order to focus on safeguards that are usable, we restrict our attention to those which have minimal adverse effect on the user experience.  
	\end{abstract}
\begin{IEEEkeywords}
	captcha, attacks, safeguards, bots
\end{IEEEkeywords}	
	\section{Introduction} \label{1}
	For almost 20 years, Completely Automated Public Turing tests to tell Computers and Humans Apart (CAPTCHAs), also known as Human Interaction Proofs (HIPs), have been used by website owners to filter out interactions from computer programs known as \textit{bots} \cite{hip,captcha}.  CAPTCHA puzzles, or tests, may take many forms, but the most commonly used involve asking the user to type in an obscured text that is intended to be difficult for bots to recognize.
	
	The type of bots that are routinely deployed against CAPTCHAs perform tasks that are unwanted by website owners.  For example, a bot might be developed to create a large number of email accounts at a free email provider, e.g.\ for subsequent use in sending spam emails \cite{naive}.  An attacker could manually create such accounts; however, they will be less productive than a bot, exemplified by the fact that we tried several publicly available bots and found that it takes less than a second for such a bot to fill out a typical registration form. Bots are widely used to perform repetitive tasks automatically and at a much higher rate than would be possible for humans \cite{detect}.  Therefore, even when stopping bots completely is not possible, it would be helpful to have means of slowing them down so that the damage they can cause is reduced.
	
	In recent years, the apparent effectiveness of methods for defeating existing CAPTCHAs has given rise to a large and growing literature on novel CAPTCHA techniques (see, for example, \cite{break, naive}).  By contrast, in this paper we focus on techniques that could stop, or at least slow down, bots by increasing the effectiveness of existing CAPTCHAs.  Several of these techniques could be useful in preventing bot traffic even if conventional CAPTCHAs are not used.  Other authors, notably Moradi and Keyvanpour \cite{review}, have looked at subsets of these safeguards, but in more restrictive contexts; in the case of Moradi and Keyvanpour they examined safeguards which can be regarded as alternatives to CAPTCHAs. 
	
	The remainder of the paper is structured as follows.  We start in sections \ref{2} and \ref{3} with background information on CAPTCHAs and the known attacks on them.  In section \ref{4} we discuss, at a high level, tactics that could be used to improve the security of CAPTCHAs, followed by descriptions of existing and novel CAPTCHA safeguards in sections \ref{5} and \ref{6}.  The safeguards are reviewed and analysed in section \ref{7}.  In sections \ref{8} and \ref{9}, we discuss potential future work and give concluding remarks.
	
	\section{CAPTCHAs} \label{2}
	CAPTCHAs are typically used by websites as an automated way of blocking interactions initiated by computer programs commonly known as bots.  These bots are used to automatically accomplish undesirable tasks such as: creating a large number of accounts (e.g.\ email accounts to be used for malicious purposes such as originating spam), manipulating online polls and creating large volumes of traffic to a website to cause a Denial of Service (DoS) \cite{captcha}.
	
	Many commonly visited websites\footnote{The CAPTCHA Usage website \url{http://trends.builtwith.com/widgets/captcha} provides continuously updated information about the current use of CAPTCHAs.} use CAPTCHAs, including YouTube, MSN, Google, Yahoo, Facebook and Twitter.  However, CAPTCHAs suffer from two main practical problems.  One is that in many cases it has been shown to be possible to program bots to solve a specific CAPTCHA.  The other is that, in practice, humans often find it challenging to solve CAPTCHAs \cite{eco}.  
	
	Existing CAPTCHAs, regardless of their type, often test one or more of the following human capabilities that are believed to be very challenging for computers to reproduce. In particular, most tests rely on the human ability to:
	
	\begin{itemize}
		\item recognize previously unseen variations of a known object, image or text,
		\item segment and recognize connected or overlapping characters and/or objects,
		\item detect anomalies through innate understanding of what is out of context \cite{invar}.
	\end{itemize}
	
	\section{Known Attacks on CAPTCHAs} \label{3}
	Many means of circumventing the protection provided by CAPTCHAs have been devised \cite{eco}, although they typically fall into one of three categories.  The first involves devising methods to automatically, and quickly, solve a particular class of CAPTCHAs, i.e.\ devising efficient software techniques for solving the problem underlying the CAPTCHA. The second involves bypassing the CAPTCHA completely, i.e.\ without solving the puzzle, typically as a result of an insecure implementation.   The third is to make humans solve CAPTCHA puzzles on behalf of bots.  We next briefly enlarge a little on each of these attack categories.
	
	\begin{enumerate}
		\item \textit{CAPTCHA-solving attacks}: Such an attack involves breaking a CAPTCHA by devising an algorithm to automatically solve the puzzle that is assumed to be unsolvable by bots.  Since most CAPTCHAs in use are visual puzzles \cite{naive}, such an attack often uses \textit{computer vision} techniques \cite{icaptcha}.  This means that the bot is able to read, or recognize, the puzzle that was wrongly assumed not to be automatically solvable.
		
		\item \textit{CAPTCHA-bypass attacks}: Attacks of this type involve a bot being able to bypass a CAPTCHA puzzle rather than solve it.  Such attacks typically rely on compromising a vulnerable implementation of a CAPTCHA, in which the server is fooled into thinking the CAPTCHA has been solved when in reality it is has been bypassed \cite{enhance}. 
		
		\item \textit{Human exploitation attacks}: In this category of attack, humans are knowingly, or unknowingly, induced to solve CAPTCHA puzzles on behalf of a bot; the human-provided solutions to the puzzles are by some means passed to the bot prior to being submitted to the website \cite{icaptcha}.  If the humans are doing so knowingly, they are typically given an incentive \cite{asirra}.  Alternatively, a human might unknowingly solve a CAPTCHA for a bot if a CAPTCHA puzzle of one website is displayed on another \cite{spam}.  
	\end{enumerate}
	\section{Improving CAPTCHA Security} \label{4}
	In the remainder of the paper, we discuss safeguards that can help stop or slow down bots.  It is important to note here that none of the safeguards should be considered as a foolproof defence against bots.  They mainly serve the purpose of reducing the chances of a bot circumventing a CAPTCHA, as well as protecting against bots that are programmed to break specific types of CAPTCHAs, e.g.\ text-based schemes.   
	
	We restrict our attention to safeguards that have a minimal effect on usability, since user acceptability is key to the success of such schemes.  That is, any safeguard that would require users to perform extra steps other than solving the CAPTCHA is not considered in this paper.  It is also important to note that CAPTCHAs could be implemented by a website owner or, alternatively, could be provided by a third-party CAPTCHA provider. For simplicity, in the remainder of this paper we assume that the CAPTCHA is provided by a third party provider, as is commonly the case.
	
	We next describe a range of possible safeguards, first those previously proposed or implemented, and then a selection of techniques we believe to be novel.  As noted above, these safeguards can be combined to create extra layers of security for a CAPTCHA.
	\section{Existing CAPTCHA Safeguards} \label{5}
	In this section, we discuss seven previously proposed CAPTCHA safeguards, namely: IP blacklisting, site keys, decoy fields, time monitoring, framing prevention, interaction detection, and pointer movement accuracy.  Two of these safeguards (described in \ref{decoy} and \ref{time}) have also been briefly discussed in \cite{review}.
	\subsection{IP Blacklisting}
	In order to protect CAPTCHAs against attack attempts from a single or a small number of IP address(es), website owners can record the IP addresses of computers interacting with their website.  Accordingly, if increased traffic from specific IP(s) is observed attempting to solve the CAPTCHA or access the website resources protected by CAPTCHA, then it could indicate that the source of interaction is a bot (this relates to a more sophisticated solution proposed by Kandula et al.\ \cite{killbot}).  IP blacklisting, i.e.\ blocking interactions arising from particularly active sites, has shown good potential in reducing spam generated by bots \cite{spam}.
	
	This method needs to be used with caution since IP addresses are sometimes shared by several computers, e.g.\ when a proxy server is in use, or the IP addresses could belong to temporarily compromised computers.  One alternative to blacklisting, which would lessen the impact on those clients falsely classified as bots, would be to increase the number or difficulty of CAPTCHA puzzles sent to such clients.  Nonetheless, since IP addresses of clients can be dynamic, it is important to remove IP addresses from the blacklist after a certain period of time.  This will help reduce the chance of blocking a client which happens to be assigned an IP address that was used by an attacker in the past.
	
	\subsection{Site Keys} \label{PKI}
	Some CAPTCHA providers create a site key (i.e.\ a unique public key) for each website making use of their CAPTCHAs.  The key helps bind the CAPTCHA to its corresponding domain.  This key is typically included in the web page's source code. For example, from some simple tests on reCAPTCHA, we discovered that the provider offers an optional feature called \textit{Domain Name Validation}, which apparently uses JavaScript to check that, when loading the page that contains the CAPTCHA, the site key used corresponds to the website it is being loaded by.  This is intended to prevent a CAPTCHA that is being used to control access to one domain from being displayed by another domain. 
	
\begin{figure}[!t]
		\centering
		\includegraphics[width=2.5in]{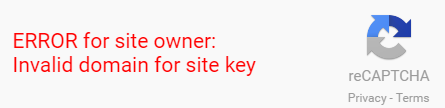}
		\caption{\textit{ No CAPTCHA reCAPTCHA} invalid domain error message}
		\label{fig:recaptcha}
		
	\end{figure}

	We tested the reuse of site keys on several widely used CAPTCHAs\footnote{The CAPTCHA Usage website http://trends.builtwith.com/widgets/captcha provides continuously updated information about the current use of CAPTCHAs.}, including reCAPTCHA, sweetCaptcha and Key CAPTCHA, by using the site key on a domain we own.  All of them seem to use this safeguard as we were unable to display CAPTCHAs originally appearing in a page for a different domain.  For example, in the case of reCAPTCHA, the page displayed the error message \textit{Error site owner invalid domain for site key} instead of the CAPTCHA, as shown in figure  \ref{fig:recaptcha}. 
	
	The reCAPTCHA scheme appears to use a somewhat more complex protection
	scheme.  The reCAPTCHA website\footnote{https://developers.google.com/recaptcha/docs/start, as accessed on 31/8/16}
	refers to the use of a website-specific key pair made up of a secret and the
	(public) site key.  Quite how these keys are related (for example, whether
	they make up a key pair for a digital signature scheme) is not made clear;
	as a result, we do not discuss the use of such key pairs further here.
	Another scheme apparently using a pair of cryptographic keys to protect
	against human exploitation attacks is described by Saklikar and Saha \cite{public},
	although it only appears to work within the context of an anti-phishing
	protection scheme; again, because of its specialised operation we do not
	consider it further here.

	\subsection{Decoy Fields} \label{decoy}
	In 2007, a blogger proposed what he called the Honeypot Captcha\footnote{ http://haacked.com/archive/2007/09/11/honeypot-captcha.aspx/} and this approach was subsequently discussed in \cite{review}.  The concept is to add an extra dummy field to an online form.  This field is made invisible by using Cascading Style Sheets (CSS) code, which hides it from users but not from bots.  The idea is to trick bots into filling out a decoy field which will not be completed by a human.  If the field is completed, then the server knows it is interacting with a bot, and the form can be rejected.  Alternatively, the decoy field could be part of the CAPTCHA puzzle instead of the form.
	
	The technique was originally proposed as a replacement for conventional CAPTCHAs requiring user interaction.  However, and like the other safeguards discussed in this paper, it seems more appropriate to consider it as a way of complementing CAPTCHAs rather than replacing them.   To make life harder for bot writers, website owners can create decoy fields using a range of techniques, e.g.\ utilizing scripting languages.
	
	\subsection{Response Time Monitoring} \label{time}
	Users will typically take longer to fill out an online form than a bot.  Indeed, if bots only worked at the speed of humans, then part of the reason for using them would disappear.  This observation motivates the operation of this safeguard, in which the website measures how long it takes for a web form to be completed and returned.  It would also be helpful to measure how quickly a user submits their input in order to detect abnormally fast responses \cite{game}.  
	
	Such a test can be implemented using a simple JavaScript timing event, and if the interaction is suspected to be from a bot because of its speed then the submission can be ignored.  Of course, the precise timing criterion used to distinguish between a bot and a human will need to depend on the type of CAPTCHA and the other inputs required, and setting this criterion correctly may need practical trials.  Moreover, to optimize the effectiveness of this safeguard it would be ideal to start measuring the time taken only once interactions have been detected rather than when the page, or CAPTCHA, is downloaded.   Such an approach will counteract bots that load pages and then wait for a short period of time before inputting the values.
	
	It is also interesting to observe that using this safeguard will either stop bots for being too quick or, alternatively, if a bot is programmed to imitate human speed, it will negatively affect one of the main objectives of using bots in the first place, i.e.\ to save time \cite{eco}.  Moreover, this safeguard would be a useful mitigation for DoS attacks, as such attacks are often meant to overwhelm a server with large volumes of traffic within a short time frame \cite{dos}.
	
	\subsection{Framing Prevention}
	Some human exploitation attacks are implemented by displaying a victim website containing a CAPTCHA on a different website using an \textit{iframe} tag or other framing technique \cite{smug}.  In this way, an attacker can take advantage of unsuspecting users to solve CAPTCHAs on behalf of a bot.  However, such an attack can be mitigated by instructing the web browser not to display a target website if it is embedded within another website \cite{click}.  This can be implemented using HTTP header tags such as x-iframe-options or Content Security Policy (CSP) \cite{busting}.
	\subsection{Interaction Detection}
	Many bots do not exhibit human-like interactions with web pages \cite{behaviour}.  This safeguard detects bots by detecting a lack of interactions with a web page that are typical of a human user \cite{mouse}.  Such interactions include: pointer movements, keystrokes and form field focus (i.e.\ placing the cursor in a field) \cite{behaviour,focus}.  This safeguard can be deployed on a full web page or restricted to the CAPTCHA. 
	\subsection{Pointer Movement Accuracy}
	Ideally, and unlike typical bots, users will not move a pointer very accurately (e.g.\ they are unlikely to move the pointer in perfectly straight lines) \cite{botcraft}.  This safeguard detects the accuracy of pointer movements to differentiate between a bot and human interaction. 
	
	This safeguard can be deployed in a range of ways, including downloading a script (e.g.\ in JavaScript) to the browser that detects when the pointer passes over certain points in a web page.  Of course, how many pointer movements need to be detected for the script to deem that a user is present will depend on the particular web page and on the CAPTCHA, and it is likely that some experimentation will be required to find appropriate parameters.
	
	It is possible that, when filling online forms, users might not use the mouse/touchpad, as they could instead use their keyboards and the \textit{Tab} key to move from one field to another.  As a result, this safeguard should ideally be restricted to the CAPTCHA interface (i.e.\ not the full web page) given that many widely used CAPTCHAs, including image-based reCAPTCHA, SweetCaptcha and KeyCAPTCHA, necessitate the use of a pointer.  
	\section{Novel CAPTCHA Safeguards} \label{6}
	In this section, we propose five new categories of safeguards that could be used to strengthen a CAPTCHA.  To the author's knowledge, none of them have previously been proposed, although we cannot rule the possibility that one or more of them may already be in use by a CAPTCHA provider.  Moreover, at least in theory, all these safeguards can be used together, as they address different aspects of detecting bot interactions.
	
	\subsection{CAPTCHA Brand Customization}
	Several CAPTCHA providers, including reCAPTCHA\footnote{\url{ https://developers.google.com/recaptcha/old/docs/customization}}, allow website owners to customize the look of  a CAPTCHA. It could be useful to take advantage of this feature so that the CAPTCHA also displays the name and/or logo of the website.  To try to ensure that this information is acted upon, the CAPTCHA could also be accompanied by a statement of the form ``Do not attempt to solve this CAPTCHA if the site name below does not match the site you are visiting". This might help alert users that they are being used to solve a CAPTCHA that belongs to a website other than the one they are visiting, i.e.\ addressing human exploitation attacks (see section \ref{2}).  Of course, providing such information would have no effect if humans are intentionally solving CAPTCHAs for bots.  
	
	This is not a particularly robust safeguard since it depends on users reading, understanding, and acting upon the information provided.  Nonetheless, if customization is enabled by the CAPTCHA provider then this should be fairly straightforward to implement.
	\subsection{Hotlink Prevention}
	Hotlinking (or inline linking) is the process of displaying an object on a web page by linking to its source \cite{inline}, and it can be used as a means to perform a human exploitation attack (see section \ref{2}).  Just like the site key safeguard described in section \ref{PKI}, hotlink prevention attempts to prevent the display of a CAPTCHA, or its corresponding page, on a website other than the one intended.  However, hotlink prevention is more straightforward to implement than the use of site key since it only requires the addition of a simple piece of code.  For example, if an Apache server is in use, the safeguard could be implemented simply by including an appropriately configured \textit{.htaccess} file.  This web server configuration file can be used to control access to server files or resources \cite{htaccess}.
	
	This safeguard is very commonly used to prevent hotlinking of images for copyright reasons and/or bandwidth abuse \cite{inline}. However, the same principle can be applied to other objects (e.g.\ web pages, audio files, etc.)\ since in all cases it will simply result in the web browser displaying an error page instead of the hotlinked object.  In the case of a CAPTCHA, any media files it uses, or even a JavaScript file that could be operating it, can be protected using such a safeguard.  Implementing such a safeguard would involve checking the HTTP referrer field of the resource request, and denying requests originating from websites other than that from which the resource request originates.
	
	There are many examples of publicly available scripts, written in a variety of scripting languages, that can be used to detect hotlinking and block the object from being displayed.  Some allow a warning message or a different message to be displayed if hotlinking is detected. 
	\subsection{Input Restriction}
	We have tested a range of publicly available bots that solve certain types of CAPTCHA, including \textit{DeCaptcher}\footnote{https://de-captcher.com}.  When the CAPTCHA requires the user to type something, the bots we tested paste the entire solution to the CAPTCHA puzzle into the web form.  By contrast, a user is unlikely to have any reason to paste the solution since they are inputting the solution to a varying CAPTCHA puzzle.  A script can be used to disable the pasting of values and, as a result, restrict input to typing only.  
	
	In order to reduce any impact on usability, we recommend that pasting is only disabled for CAPTCHA puzzle solutions and not for form inputs, since some users might be using their browser's autofill feature or might choose to paste, rather than type, certain values such as email addresses, phone numbers, etc.  Of course global disabling of pasting has some advantages, since it will help stop some bots, or at least slow them down, as all field values will need to be typed one character at a time.  Ultimately, this is a usability issue and website owners must decide whether to disable pasting of values for the entire web form or just for the CAPTCHA puzzles, depending on the nature of the page and their own risk assessments.  
	
	\subsection{Device  Fingerprinting}
	Device fingerprinting techniques, as discussed for example by Eckersley \cite{unique}, are extremely effective in uniquely identifying a single client.  They typically involve a website downloading JavaScript which interrogates the browser to learn, for example, which fonts are enabled.  The combination of information available to the script is, apparently, sufficient to uniquely identify a client platform with high probability, independently of the IP address.  Indeed, the technique is so effective that there are a number of commercial providers of device fingerprinting services\footnote{see for example https://www.cybersource.com/}.
	A bot will typically operate from a single server, or a small number of servers, and so if device fingerprinting reveals a significant number of requests all originating from the same device then this could be a strong indication of a bot.  That is, fingerprinting could be used to blacklist potential bot hosts, analogously to IP blacklisting.
	\subsection{Switching between CAPTCHAs}
	Typically, bots are programmed to attack specific CAPTCHAs \cite{reverse}. This safeguard works by switching between multiple CAPTCHA types.  For example, if a website uses reCAPTCHA and a bot is programmed to defeat it, then the website would be virtually defenceless against that bot.  However, if the website switches between different CAPTCHAs, then the website would be vulnerable to that bot only when an instance of reCAPTCHA is used.  This would significantly reduce the number of successful attacks by such CAPTCHA-specific bots.  
	
	Implementation would typically involve automatically switching between CAPTCHAs from a range of CAPTCHA providers.  However, using in-house CAPTCHAs is also possible, as long as there is a significant difference between one CAPTCHA and another.  The strength of such a safeguard is clearly dependent on a variety of factors including: the strength of CAPTCHAs deployed, the total number of CAPTCHA types deployed and the frequency of switching.  Moreover, the effect on usability will vary depending on the selected CAPTCHAs.

	\section{Safeguard Effectiveness} \label{7}
	We next provide an evaluation of the effectiveness of the 12 safeguards introduced in the previous two sections.  We start by discussing the criteria for performing the evaluation, and then examine each safeguard in turn.  A summary table is provided in section \ref{sum}.

	\subsection{Evaluation Criteria}
	In order to understand the effectiveness of safeguards we first need to decide how to perform the evaluation.  We have chosen to use the following four evaluation criteria, for the reasons specified.
	\begin{itemize}
		\item \textit{Protection offered:} In section \label{3} we identified three general classes of attack against CAPTCHAs. Some safeguards only help protect against certain classes of attack, and so for each technique we evaluate which of the three main attack types it can help mitigate.
		
		\item \textit{Implementation possibilities:} Since such an approach is widely used, we need to consider the possibility that the CAPTCHAs are provided by a third party, independent of the website the CAPTCHAs are being used to protect. All the safeguards we discuss can be implemented with the support of the CAPTCHA provider, but only some of them can be deployed by a website independently of the provider. Clearly, allowing the latter is potentially advantageous in practice since website owners will be able to improve CAPTCHA security for themselves, instead of depending solely on CAPTCHA providers.  Also, there are inherent advantages in site-specific implementations of attack mitigations, since ``off the shelf" bots may be ineffective against them.
		\item \textit{Effect on usability:} Some safeguards might, in some circumstances, negatively affect the user experience, albeit in a minor way.  Others operate in such a way that they are most unlikely to have any effect on the user experience.
		
		\item \textit{Simplicity of implementation:} All safeguards require the server to validate the interaction in some way.  In some cases, the validation process can be implemented easily, e.g.\ by adding simple HTML or CSS code.  Others may be a little more complex to implement, e.g.\ requiring the use of client-side scripting.  Yet others may require quite complex functionality, e.g.\ involving server-side scripting/programming.  This criterion measures the expected difficulty of implementing a safeguard by a website owner.  In practice, the difficulty of implementation will vary depending on the system being protected and the detailed approach selected to implement the safeguard. 
	\end{itemize}

	\subsection{Evaluation}
	We next use the criteria proposed immediately above to evaluate the effectiveness of each of the safeguards.  We consider the 12 safeguards in the order in which they were introduced in sections \ref{5} and \ref{6}.
	\begin{itemize}
		
		\item IP Blacklisting: This safeguard could be used to help protect against all three types of CAPTCHA attacks, as it aims at blocking any IP address that generates suspiciously large volumes of traffic to a website.  This is ideally implemented by a website owner as it would specifically target attackers of their website.  Unfortunately, this safeguard carries with it the risk of blocking legitimate users who happen to share a public IP address with an attacker, are using a proxy server, or who are using a compromised machine.  
		
		An IP blacklisting service is available from many web hosting providers\footnote{See for example, HostGator or GoDaddy}.   However, if this option is not available from a hosting provider or no hosting provider is involved, then this safeguard is relatively complex to implement.  This is because its implementation would require a means of monitoring the source IP addresses of visits to a website as well as providing server-side code to block particularly active IP addresses (e.g.\ by configuring the Linux native firewall, \textit{iptables}).
		
		\item Site Keys: This safeguard prevents a CAPTCHA from being displayed on a website other than the one intended, thereby mitigating human exploitation attacks.  This safeguard can only be implemented by a CAPTCHA provider, as it requires the generation of a key that is held by both the website and the CAPTCHA provider.  This safeguard is unlikely to affect legitimate users of a website in any way, as it is designed only to prevent the display of a CAPTCHA on an unauthorised website.  Once a key is provided to the website owner, implementing it may be as simple as adding a few lines of code\footnote{This is described on the help page: https://developers.google.com/recaptcha/docs/display}.
		
		\item Decoy Fields: This safeguard helps protect against bots that attempt to automatically fill form fields and submit CAPTCHA puzzle solutions.  It can be implemented by the website owner, independently of the CAPTCHA provider, by adding a decoy field to its online form and rejecting any submission that has this field filled in.  Since the decoy field is invisible, it is unlikely that, in practice, a user would be affected by it. However, if the user is employing an autofill program, and this program entered values in the decoy field just like a bot, a false alarm could result, although such an event seems unlikely.  Implementing this safeguard simply requires the addition of an extra field in a form using HTML, and using CSS to hide it\footnote{This was demonstrated by the blogger who first proposed this approach -- see: http://haacked.com/archive/2007/09/11/honeypot-captcha.aspx/}.
		
		\item Response Time Monitoring: This safeguard protects against bots which complete forms or solve CAPTCHAs significantly faster than humans would.  Accordingly, it defends against CAPTCHA-solving attacks.  It can be implemented by the CAPTCHA provider if monitoring is restricted to the time needed to solve the puzzle.  Alternatively, if the safeguard is implemented by the website owner, e.g.\ in the form of JavaScript downloaded to the browser, it could monitor the time required to complete a form and solve the CAPTCHA.  Just like the previous safeguard, a user might be negatively affected if using an autofill program, as it might complete a form faster than a user would.   One of the ways this safeguard can be implemented is through the use of client-side scripting to calculate the time elapsed from the moment input is detected.  The rate of CAPTCHA/form completion can be calculated from the differences between timestamps of monitored events.

			\begin{center}
				\begin{table*}
					 \centering
					\caption{CAPTCHA Safeguards Comparison}\label{tab1}
					\begin{tabular}{ll m{9em}  ll}
						\hline\noalign{\smallskip}
						Criteria / CAPTCHA & \textbf{Protects Against?}  & \textbf{Implementable By Website owner?} & \textbf{Might Affect Usability?} & \textbf{Simplicity}\\  
						\hline
						IP Blacklisting & All Attacks & Yes  & Yes & Complex\\ 
						\hline
						Site Keys & Human exploitation & No &  No & Simple\\				
						\hline
						Decoy Fields & CAPTCHA-solving & Yes  & Yes & Simple\\
						\hline
						Response Time Monitoring & CAPTCHA-solving & Yes  & Yes & Moderate\\
						\hline
						Framing Prevention & Human exloitation  & Yes  & No & Simple\\
						\hline
						Interaction Detection & CAPTCHA-solving  & Yes  & No & Moderate\\
						
						\hline
						Pointer Movement Accuracy & CAPTCHA-solving  & Yes  & Yes & Complex\\
						\hline
						
						Brand Customization & Human exloitation & No  & No & Simple\\
						\hline
						Hotlink Prevention & Human exloitation & Might  & No & Simple\\  
						\hline
						Input Restriction & CAPTCHA-solving & Yes  &  Yes & Moderate\\
						\hline
						Device Fingerprinting & All Attacks  & Yes  & Yes & Complex\\
						\hline
						Switching between CAPTCHAs & CAPTCHA-solving  & Yes  & Yes & Complex\\
						\hline

					\end{tabular}
				\end{table*}
			\end{center}

		\item Framing Prevention: This safeguard prevents users from interacting with a website that is embedded within another.  This will prevent users from being tricked into solving CAPTCHAs of websites other than they are visiting.  It can be implemented independently from CAPTCHA providers as it involves configuring HTTP headers on the website owner server side.  This safeguard is not expected to have any impact on legitimate users.  Framing prevention is simple to implement, e.g.\ by the addition of a few lines of code to the server's configuration file.
		\item Interaction Detection: This safeguard detects the presence of human-like interactions, such as pointer movements and keystrokes.  It therefore protects against bots attempting to solve CAPTCHAs but that are not programmed to mimic typical user interactions.  A website owner could deploy such a safeguard independently of the CAPTCHA provider.    It is unlikely to affect users whether they are using a mouse/touchpad or solely depending on a keyboard for interacting with the website, as any type of human activity can be made to trigger one of the interaction events and so be deemed as indicative of a legitimate user.  This could, for example, be implemented using \textit{JavaScript Events}\footnote{see \url{http://www.w3schools.com/js/js_events.asp}}, in which case a CAPTCHA solution would only be accepted if one of these events is triggered. 
		\item Pointer Movement Accuracy: This safeguard helps a server distinguish between a user and bot.  A website owner can implement it independently of the CAPTCHA provider as it involves triggers similar to those described under \textit{Interaction Detection}.  However, in this case they are used to measure the accuracy of movements rather than their presence.  If a form or a CAPTCHA cannot be completed without the use of mouse movements, then this safeguard will successfully detect a bot if its mouse movements are too precise.  However, depending on the accuracy of the detection system, it might also cause a false positive.  This safeguard could be implemented using inputs from the JavaScript \textit{onmouseover} feature that could be implemented at various locations on a web page or in the CAPTCHA itself.  Nevertheless, it would probably be quite complex to program the number and location of triggering events necessary to accurately identify a bot.
		\item Brand Customization: This safeguard will alert unsuspecting users that they are being fooled into solving a CAPTCHA puzzle of another website.  This safeguard can only be implemented if the CAPTCHA provider provides a customization option to the website owner.  The addition of a web logo, and perhaps a small amount of warning text, is unlikely to have a significant adverse effect on users.  Since this is dependent on the CAPTCHA provider, implementing it would not typically require a website owner to do any more than take a few simple steps to display the customized CAPTCHA.
		\item Hotlink Prevention: This safeguard helps protect against human exploitation attacks, as it helps prevent a CAPTCHA from being redisplayed on a different website.  If a CAPTCHA provider requires files (e.g.\ JavaScript files) to be stored on the client server, then this safeguard can be implemented independently of the CAPTCHA provider. Just like \textit{Site Keys}, this aims solely at preventing the display of a CAPTCHA on a website other than that for which it was intended, and so is unlikely to affect the user experience in any way.  As previously described, this safeguard can be implemented very simply by configuring the .htaccess files on the web server. 
		\item Input Restriction: This safeguard helps protect against CAPTCHA-solving bots by preventing pasting of a CAPTCHA solution or even pasting of values into fields in an online form.  Depending on the type of CAPTCHA, this could be implemented by the CAPTCHA provider or the website owner.  The website owner can implement it anywhere in the page that contains the CAPTCHA, while the CAPTCHA provider can incorporate it in the CAPTCHA itself as long as it has a field that requires typing of values.  
		
		It might affect usability if pasting is disabled for the whole page as some users might want to utilise pasting, e.g.\ as part of autofill.  However, if pasting is disabled only for the solution to a CAPTCHA, then a user would not typically be affected by it as there is no obvious reason to paste the solution to the puzzle.  This safeguard could be implemented in a variety of ways, e.g.\ by using JavaScript code that only allows the input of one character at a time in a given field.
		
		\item Device Fingerprinting: This safeguard serves the same purpose as IP blacklisting.  Nevertheless, it has the advantage of identifying specific devices regardless of whether they use changing IP addresses or they share IP addresses with other benign devices.  However, it is rather more complicated to implement as it requires the identification of device fingerprints by collecting a range of information about the client platform, typically including the OS and web browser.   Device fingerprinting can be implemented by the website owner or outsourced to a device fingerprinting service provider.  This safeguard aims at blocking interactions with suspicious web clients or at least prompting the use of extra, perhaps even more complex, CAPTCHA puzzles.  Legitimate users might be affected if they were mistakenly identified as bots and as a result blocked from interacting with the website or challenged with more CAPTCHA puzzles.
		\item Switching between CAPTCHAs:  This safeguard reduces the success chances of bots programmed to solve specific CAPTCHAs.  It can only be implemented by a website owner, as it would typically involve deploying CAPTCHAs from a range of providers.  It might affect usability as the user experience would vary depending on the difficulty of the CAPTCHA presented.  Implementing this safeguard is expected to be quite complex, as it would require programming a website to automatically switch between CAPTCHAs as well as deal with different types of response depending on the CAPTCHA provider.
		
	\end{itemize}
	Large scale practical trials are required to complete the comparison of the performance and effectiveness of the safeguards discussed in this paper. Trials of specific safeguards are planned as future work (see section \ref{8}).

	\subsection{Summary of Evaluation} \label{sum}
	The results of our evaluation are summarised in Table \ref{tab1}. Column 2 of the table indicates which attack (as discussed in section \ref{3}) is mitigated by a given safeguard.  Column 3 indicates whether or not a safeguard can be independently implemented by a website owner.  Column 4 indicates whether or not a safeguard could have an effect on usability.  In column 5 the simplicity of implementing a safeguard is summarised as one of: HTML (simple), client-side scripting (moderate) and server-side scripting/programming (complex).

	\section{Future Work} \label{8}
	Given their growing ubiquity, additional safeguards could usefully be developed to take advantage of particular properties of touchscreen devices such as smartphones.  For example, very accurate touches could be deemed suspicious, and may possibly be indicative of a bot interaction.  Another possibility would be to request users to stroke their fingers across certain paths or points.  This is a similar approach to Android's pattern phone lock security, but in this case it would be used by a website to prove human interaction.  Additionally, sensors commonly found in modern smartphones such as motion sensors, gyroscope, etc.\, could be utilised as evidence of human interaction.
	
	Ultimately, an invisible CAPTCHA would be ideal, as it would be hidden from users and not require any specific actions by them; hence it would not create any usability issues that might frustrate users.  Such a CAPTCHA would ideally work in the background, and use information derived from normal human interactions with the website to confirm they are human.  It therefore seems reasonable to focus future research efforts on developing such non-intrusive human/computer distinguishers.     
	
	The more difficult CAPTCHA-solving becomes for bots, the more attackers are likely to move towards the use of human CAPTCHA solvers.  We believe that one of the biggest challenges in the future will be to stop humans doing the dirty work on behalf of bots.  Addressing this will require designing CAPTCHAs that not only differentiate between human users and computers, but that also have the ability to differentiate between genuine and non-genuine users.
	\section{Conclusion} \label{9}
	In general, layering is better than depending on a single layer of protection (the defence in depth principle).  This is why depending on a CAPTCHA puzzle alone as the sole means of filtering out bot interactions is not necessarily the most effective approach.  A bot is programmed to successfully attack a specific type of CAPTCHA and, as a result, a website that uses such a CAPTCHA as its sole defence will immediately be vulnerable to attacks by this bot.  Since past experience suggests that the development of a CAPTCHA-defeating technique is always possible (or even likely), supplementary safeguards such as those discussed in this paper are a vitally important weapon in the fight against bots.

	\section*{Acknowledgments}
	I would like to thank Professor Chris Mitchell for his guidance, encouragement and advice.
	
	\bibliographystyle{plain} 
	\bibliography{impcaptcha}
	
\end{document}